\documentclass[a4,12pt]{revtex4}
\usepackage{amssymb}
\usepackage{amsmath}
\usepackage{graphicx}
\begin{document}
\thispagestyle{empty}

\title{A $^{10}$B-based neutron detector with stacked Multiwire Proportional Counters and  macrostructured cathodes}

\author{I. Stefanescu$^1$,Y. Abdullahi$^1$, J. Birch$^2$, I. Defendi$^1$, \\
R. Hall-Wilton$^3$, C. H\"oglund$^{2,3}$, L. Hultman$^2$, M. Zee$^1$, K. Zeitelhack$^1$}    

\affiliation{$^1$Forschungs-Neutronenquelle Heinz Maier-Leibnitz, Technische Universit\"at M\"unchen, D-85748 Garching, Germany}
\affiliation{$^2$Department of Physics, Chemistry and Biology (IFM), Thin Film Physics Division, Link\"oping University, SE-581 83 Link\"oping, Sweden}
\affiliation{$^3$European Spallation Source ESS-AB, P.O. Box 176, SE-221 00 Lund, Sweden}

\begin{abstract}
We present the results of the measurements of the detection efficiency for a 4.7 \r{A} neutron beam incident upon a detector incorporating a stack of up to five MultiWire Proportional Counters (MWPC)  with Boron-coated cathodes. The cathodes were made of Aluminum and had a surface exhibiting millimeter-deep V-shaped grooves of 45$^\circ$, upon which the thin Boron film was deposited by DC magnetron sputtering. The incident neutrons interacting with the converter layer deposited on the sidewalls of the grooves have a higher capture probability, owing to the larger effective absorption film thickness. This leads to a higher overall detection efficiency for the grooved cathode when compared to a cathode  with a flat surface. Both the experimental results and the predictions of the GEANT4 model suggests that a 5-counter detector stack with coated grooved cathodes has the same efficiency as a 7-counter stack with flat cathodes. The reduction in the number of counters in the stack without altering the detection efficiency  will prove highly beneficial for large-area position-sensitive detectors for neutron scattering applications, for which the cost-effective manufacturing of the detector and associated readout electronics is an important objective. The proposed detector concept could be a technological option for one of the new chopper spectrometers and other instruments planned to be built at the future European Spallation Source in Sweden. These results with macrostructured cathodes generally apply not just to MWPCs but to other gaseous detectors as well.
\end{abstract}

\maketitle
\date
\newpage

\section{Introduction}
\label{intro}


In a recent paper by this collaboration we reported on the results of the theoretical and experimental studies of a novel design for the parallel cathode planes of a Multiwire Proportional Counter (MWPC) \cite{Ste01}. The proposed design consisted of a 3D regular pattern in the form of millimeter-deep grooves with an opening angle of $\alpha$=45$^\circ$, upon which a thin film of  $^{10}$B$_4$C is deposited. Filled with a commercial counting gas, the MWPC with Boron-coated cathodes becomes sensitive to thermal and slow neutrons. The groove structure helps increasing the detection efficiency over that of a layer deposited on a flat substrate with the same active area, because the Boron converter deposited on the sidewalls of the grooves has a higher probability not only to capture and convert the incoming neutron, but also release the reaction products \cite{Gre03}.  In  Ref.  \cite{Ste01} we employed the GEANT4 package \cite{geant} to study the dependence of the detection efficiency with the size of the groove. The model predictions for the optimal height of the groove were tested in measurements with monoenergetic neutron beam provided by the FRM-II research reactor in Munich \cite{frm2}.  

Because the thin Boron film was applied on a corrugated surface, the conformality of the coating was also carefully studied in our previous work \cite{Ste01}. The experimental efficiency values and absorption of the neutron beam in the Boron layer, obtained by scanning the surface of the grooves with a collimated neutron beam, were compared to the  results of the GEANT4 simulations in which two different coating scenarios were assumed. In the first scenario, the coating was modeled uniformly distributed onto the surface of the groove. In that case, the effective neutron absorption film thickness is the same anywhere on the surface of the coated macrostructured plate. In the second scenario, the coating was assumed to be preferentially distributed on the upper sides of the groove, with the thickness of the coating decreasing with decreasing the distance from the bottom of the groove. The GEANT4 calculation of the efficiency of coated plates with grooves of various heights suggested that both coating scenarios give similar results for grooves not deeper than $\sim$2.5 mm. As the GEANT4 model does not include the generation of the low energy electrons in the counting gas and their drift in the electric field applied at the anode wires,  the influence of the macrostructured surface on the charge collection efficiency could not be studied theoretically. However, the experimental data presented in Fig. 19 of Ref. \cite{Ste01} was found to agree very well with the calculation without the electric field phenomena, which could be regarded as an indication that the charge collection efficiency is little affected by the presence of the small grooves created on the surface of the cathodes.         
     
\begin{figure*}[ht]
\centering
\includegraphics[scale=0.7]{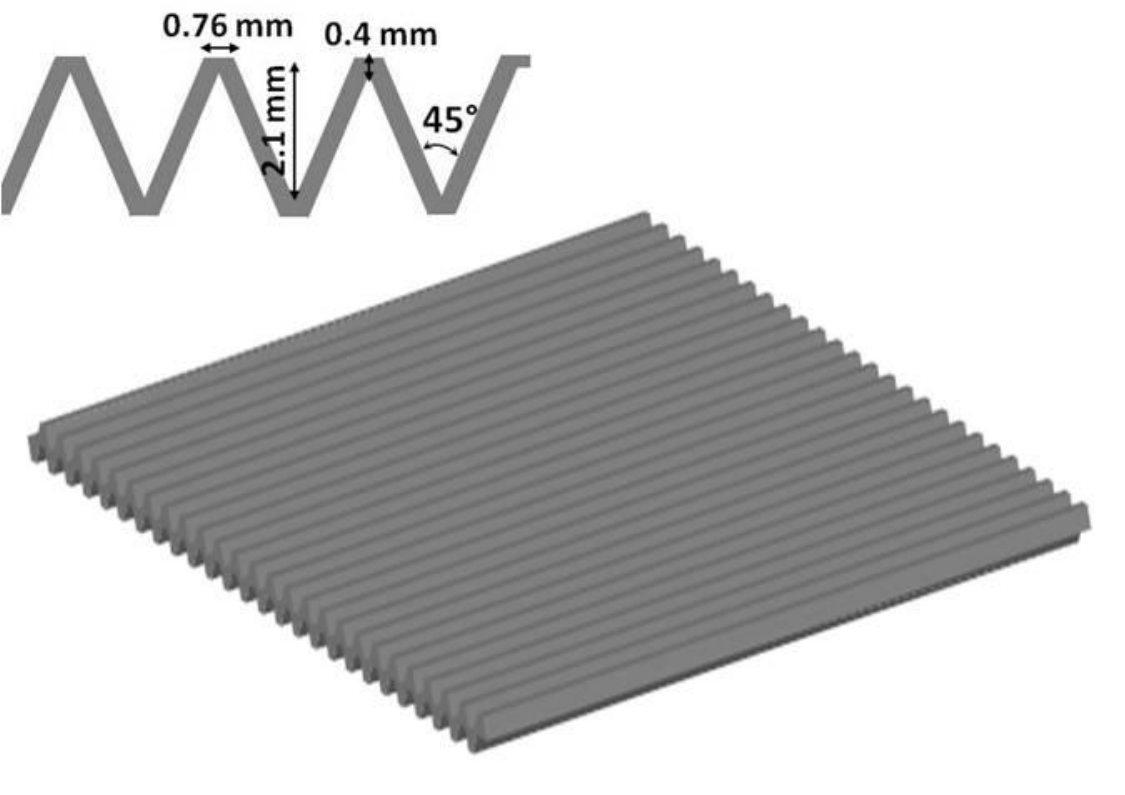}
\caption{Geometry and shape parameters for the macrostructured cathode used for the efficiency measurements performed in this work.}  
\label{des}
\end{figure*}

In this paper we present and discuss the results of the investigations on the detection efficiency of  single and multiple Boron layers coated on the surface of thin, flat plates made of Aluminum, as well as surfaces exhibiting 2.1 mm deep grooves in double-sided format. The exact geometry of the groove is illustrated in the inset of  Fig. \ref{des}.  The opening angle of 45$^\circ$ selected for the groove was the best compromise between easy and fast manufacturing and efficiency performance.  Thin films of $^{10}$B$_4$C with various thicknesses were deposited onto the Aluminum substrates by employing the technique of DC magnetron sputtering, developed in the Thin Film Physics Division of the Link\"oping University in Sweden \cite{Hog12}.  The experimental efficiencies were measured with a neutron beam of 4.7 \r{A} delivered by the TREFF instrument in operation at FRM-II \cite{Trf}.  The measured data are compared to the results of the simulations performed with the GEANT4 model for setups which resembled closely the experimental arrangements. 

\section{GEANT4 simulations for the detection efficiencies}
\label{g4}

The modeling of the detection efficiencies for the two types of cathodes was carried out in GEANT4 version 4.9.6 \cite{geant}. A detailed description of the modeling framework is given in Ref. \cite{Ste01} and it will not be repeated here. The theoretical setup incorporated the detector housing made of Aluminum, the counting gas (Ar/CO$_2$, 70/30), and the $^{10}$B$_4$C coatings placed in close contact with the Aluminum-substrates. The calculations shown here were performed  by assuming that the Boron converter is preferentially deposited on the upper part of the sidewall of the groove, with the thickness of the coating decreasing with decreasing the distance from the bottom (cf. Fig. 3a in Ref. \cite{Ste01}). The energy deposited in gas was recorded in a histogram for each of the two reaction products emitted in the direction of the gas with a kinetic energy that was high enough to make the escape from the converter possible. The theoretical efficiencies were determined from the integral of the energy histograms divided by the number of incident neutrons.       

\section{Experimental set-up}
\label{lay_det}

\begin{figure*}[ht]
\centering
\includegraphics[scale=0.6]{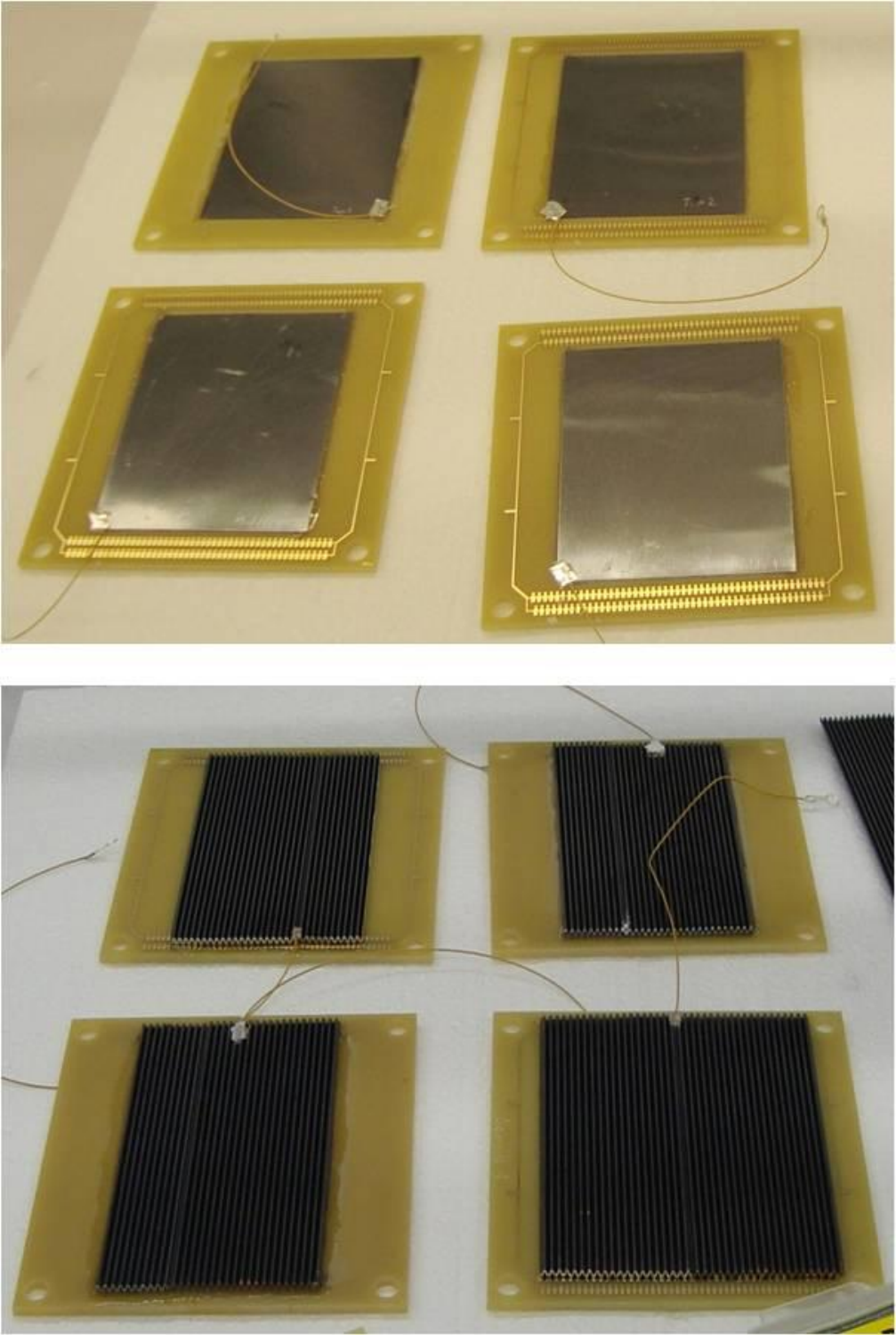}
\caption{Photographs of the coated cathodes used in the efficiency measurements reported in this work. Top:  $^{10}$B$_4$C coatings of 0.95 $\mu$m deposited by DC magnetron sputtering on Aluminum-substrates with a flat surface. Bottom: coatings of 1.1 $\mu$m deposited on Aluminum-substrates exhibiting 2.1 mm deep grooves with an opening angle of 45$^\circ$. }  
\label{lay}
\end{figure*}

The Boron layers were deposited on substrates made of Al-Mg-3 alloy (Aluminum-5754). Several plates with a flat surface and a size of 10x12x0.5 cm$^3$  were coated on one or both sides with 0.95, 2.45 and 3 $\mu$m of enriched (98 at. $\%$) $^{10}$B$_4$C. The coated plates were glued onto frames made of PCB material with an area of 13$\times$13 cm$^2$, as shown in Fig. \ref{lay}.  The grooved plates were fabricated by extrusion on Al-Mg-Si-0.5 (AW6060T-6) alloy by MIFA Aluminium BV \cite{Mifa}, and coated with 1.1, 1.6, 1.8, and 2.95 $\mu$m of $^{10}$B$_4$C. The thickness of the Boron layers deposited on the flat plates and tops of the grooves was determined by analyzing the thickness of the layers deposited on small Silicon samples with smooth surfaces which were mounted in the center of the deposition chamber and coated at the same time with the various Aluminum-substrates \cite{Hog12}.  

The test detector used to assess the efficiency of the Boron layers was described in Ref. \cite{Ste01}. The detector is made of Aluminum and has the form of a square parallelepiped of size 19.8$\times$19.8$\times$6.8 cm$^3$.  It can house up to four macrostructured plates mounted in two independent MWPCs. The coated plates were fixed on the detector lid with the help of four cylindrical rods. The high voltage was supplied to the sense wires via SHV feedthroughs mounted in the detector lid.  The wire planes used to amplify and collect the signal were fabricated by soldering tungsten wires with a diameter of 20 $\mu$m and 5 mm pitch onto PCB frames.  

\section{Single-layers efficiencies}
\label{single_l}

The Aluminum-plates supporting the Boron converter to be investigated for efficiency were mounted one by one in the counter closest to the entrance window for the neutron beam. The wire plane was positioned at 6.5 mm distance from the coated plate. The MPWC was completed with a bare Aluminum-plate, identical in size and shape with that used as a substrate material for the Boron layer.  During the measurements the detector was continuously flushed with the Ar/CO$_2$ (70/30) gas mixture. 

The readout of the signal was performed with an MRS2000 preamplifier from MESYTEC \cite{Mesy}, which also incorporates a square shaper with a shaping time of 3.5 $\mu$s. The signal from the preamplifier was fed directly into a 2K multichannel analyzer, which had a dynamic range of 5 V. Positive high voltage was applied on the wires, and the gas gain was adjusted such that the endpoint of the pulse-height spectra, corresponding to the $\alpha$-particle energy of 1.87 MeV, falls around channel 1900.  

\begin{figure*}[ht]
\centering
\includegraphics[scale=0.55]{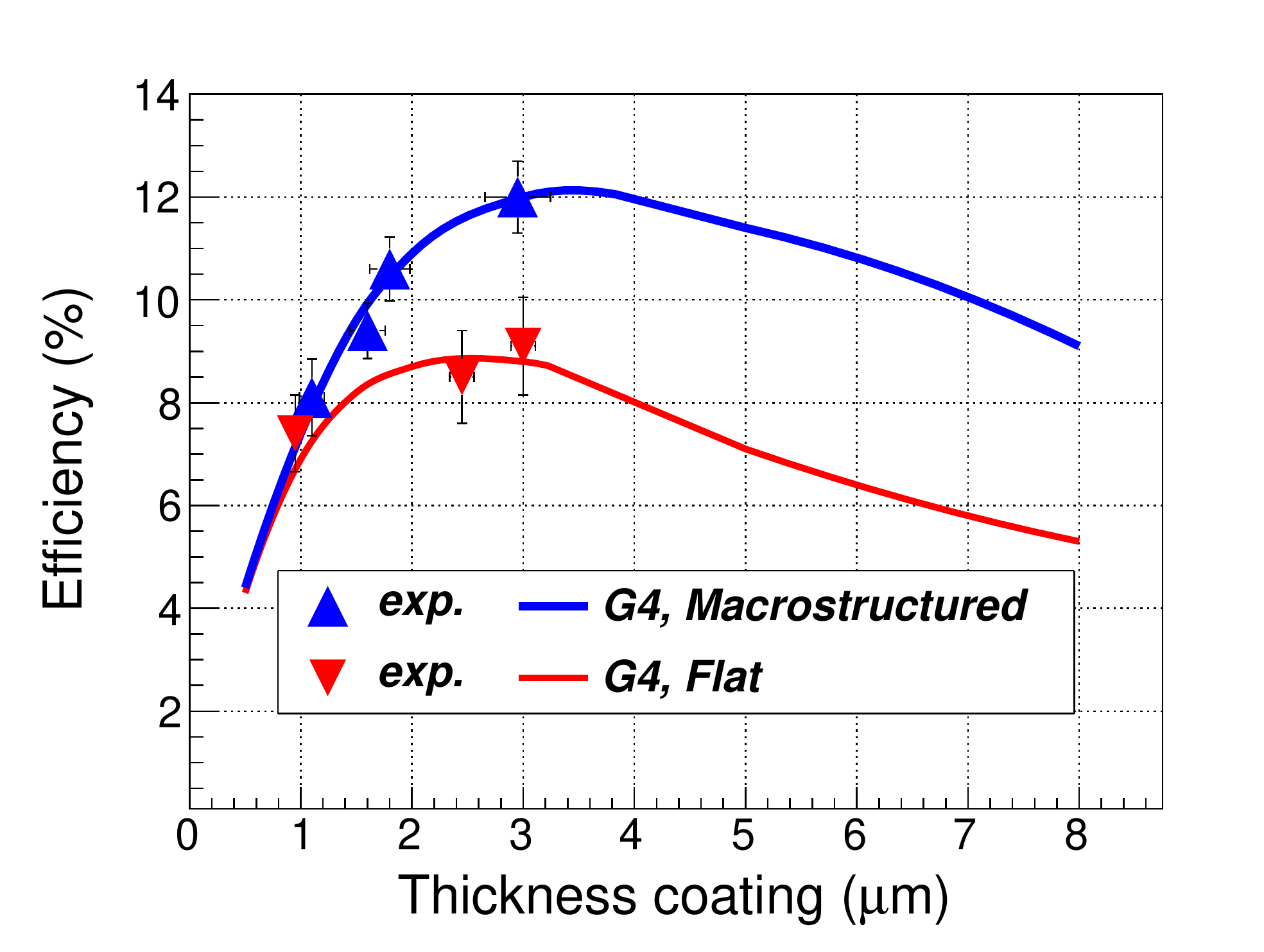}
\caption{Calculated and measured efficiencies for single layers of (98 at. $\%$)  $^{10}$B$_4$C with various thicknesses. The blue symbols and continuous  blue line  were obtained for a coated surface exhibiting 2.1 mm deep grooves of 45$^\circ$. The red symbols and continuous red line correspond to the experimental and simulated results obtained for coatings on a flat Aluminum-substrate, respectively. The efficiency is represented as a function of the thickness of the coating measured by cross-sectional SEM on the small Silicon surfaces attached to the samples mounted close to the center of the sputtering targets. The horizontal error bars  account for the variation in thickness with mounting position along the target, as discussed in Ref.  \cite{Hog12}.}  
\label{eff_s}
\end{figure*}

The neutron beam used in all investigations reported here had an energy of 4.7 \r{A}. The beam was collimated down to a spot of ca. 3 mm$^2$ with the help of  slits mounted between the monochromator and the detector table. The beam intensity was monitored at all times during the measurements by means of a fission chamber mounted before the last collimation slit. All efficiency measurements were made in the same geometry and experimental conditions. Several times during the beamtime, the detector was removed from the table and a calibrated 10-bar, $^3$He-filled, 1-inch tube was placed instead. The measured efficiency for the $^3$He counter was 96.5(3)$\%$. The efficiency of the Boron layers was determined by normalizing the number of counts observed in the corresponding pulse-height spectra to the number of neutrons detected by the reference $^3$He detector. 

The results of the efficiency measurements for the single Boron layers on flat and macrostructured Aluminum-substrates are shown in Figure \ref{eff_s}. The full red and blue lines correspond to the results of the GEANT4 simulations. The data  for the macrostructured layers  is represented as a function of the thickness of the Boron layer determined in the analysis of the Silicon sample with smooth surface that was coated in the same deposition run. The value determined for the thickness of the coating on Silicon corresponds to that of the  layer deposited on the flat portion joining the adjacent grooves, see Fig. \ref{des}. 

The experimental results presented in Fig. \ref{eff_s} fall on the respective theoretical lines, confirming the theoretical prediction that the grooving of the substrate material leads to an increase in efficiency by almost 40$\%$ with respect to a layer deposited on a flat substrate. The main contribution to the overall efficiency is given by the grooves, which occupy over 80$\%$ of the active surface of the cathode. As the thickness of the film coated on the sidewalls of the grooves is always lower than that deposited on the flat tops, the optimum thickness of the converter is larger for the macrostructured surface than for a flat surface of the same area.  Thus, for a single Boron layer deposited on a plate with the surface covered by 2.1 mm deep grooves of 45$^\circ$  the optimum thickness is $\sim$3.5 $\mu$m, as opposed to 2.5 $\mu$m determined for a flat layer, see Fig. \ref{eff_s}. Both the experimental and calculated efficiencies  were extracted from the measured and simulated pulse-height spectra  with the threshold for valid neutron events set at 120 keV.  The choice of this energy value to discriminate between the neutron signal and  $\gamma$-background was made based on the results of the experimental and theoretical investigations on the response of our small test detector to the $\gamma$-field emitted by a moderated $^{252}$Cf-source, see Ref. \cite{Ste01}.  

\begin{figure*}[ht]
\centering
\includegraphics[scale=0.5]{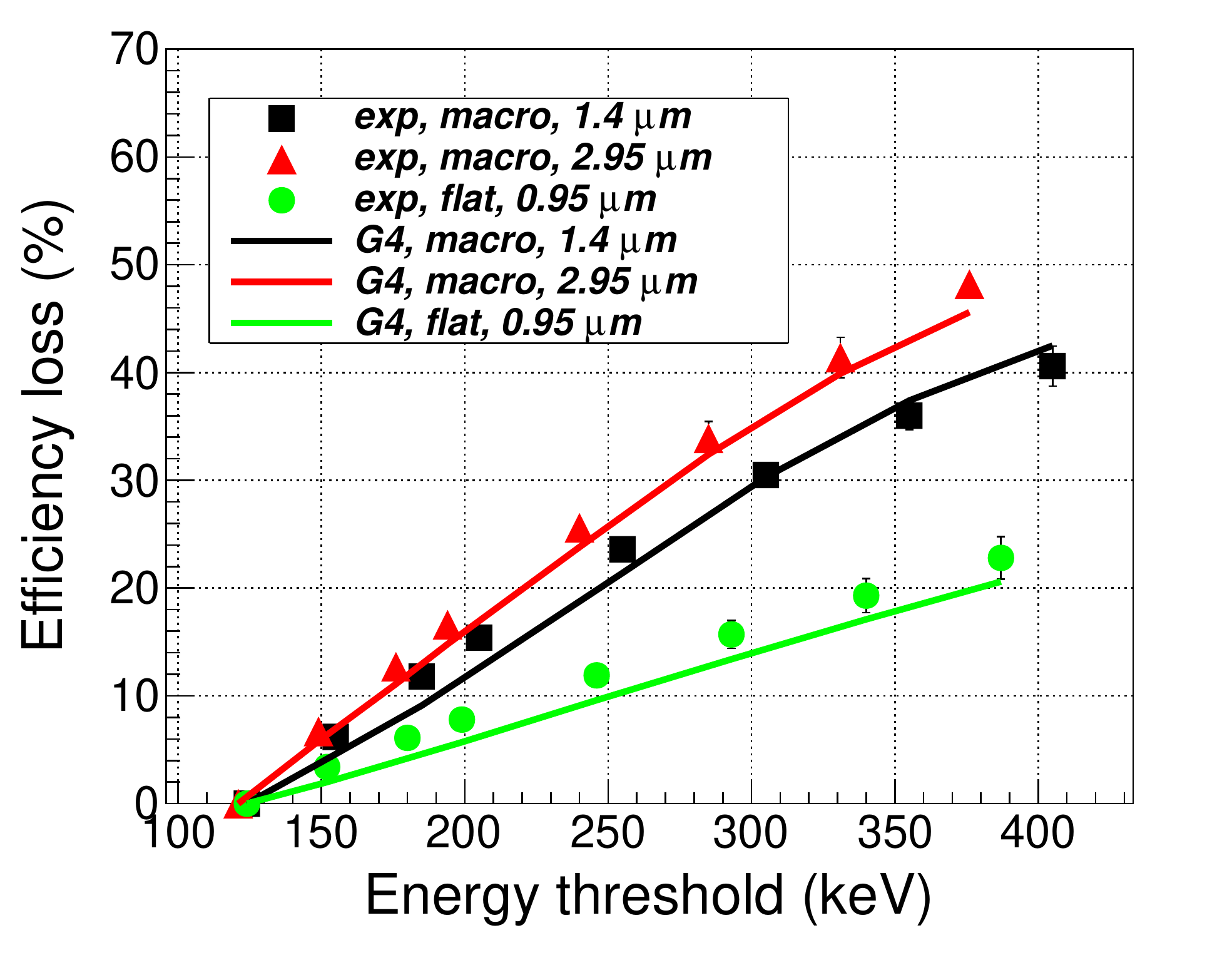}
\caption{Efficiency loss caused by increasing the n-$\gamma$ discriminator threshold from 120 keV to $\sim$400 keV for a macrostructured cathode coated with 1.1 (black squares) and 2.95 $\mu$m (red triangles), and a flat cathode coated with 0.95 $\mu$m of $^{10}$B$_4$C (green bullets). The full lines correspond to the results of the  GEANT4 calculations. The data from each set was normalized to the number of counts recorded in the corresponding pulse-height spectrum above 120 keV. }  
\label{thr}
\end{figure*}

Figure \ref{thr} shows the calculated and measured loss in efficiency caused by increasing the $\gamma$-rejection threshold from 120 keV to $\sim$400 keV. The summed number of neutron events corresponding to each threshold value considered in the analysis was normalized to that obtained with the threshold set at 120 keV.  The experimental data indicates that by rising the n-$\gamma$ discriminator threshold from 120 keV to 200 keV, causes an additional loss of $\sim$16$\%$ in the detection efficiency for the macrostructured cathodes and $\sim$8$\%$ for the cathodes with a flat surface. The plot also shows that the efficiency loss due to n-$\gamma$ threshold depends on the thickness of the Boron coating. For thin coatings the energy lost by the reaction products in the converter itself decreases and the charged ions reach the gas phase with higher kinetic energies. The events in the energy deposition spectrum  are in this case more uniformly  distributed over the whole dynamic range, which leads to a smaller overall loss in detection efficiency with increasing threshold.  For all three investigated cases, the  calculation agrees well  with  the values observed experimentally.       

\section{Multilayers efficiencies}
\label{mult_l}

\begin{figure*}[ht]
\centering
\includegraphics[scale=0.6]{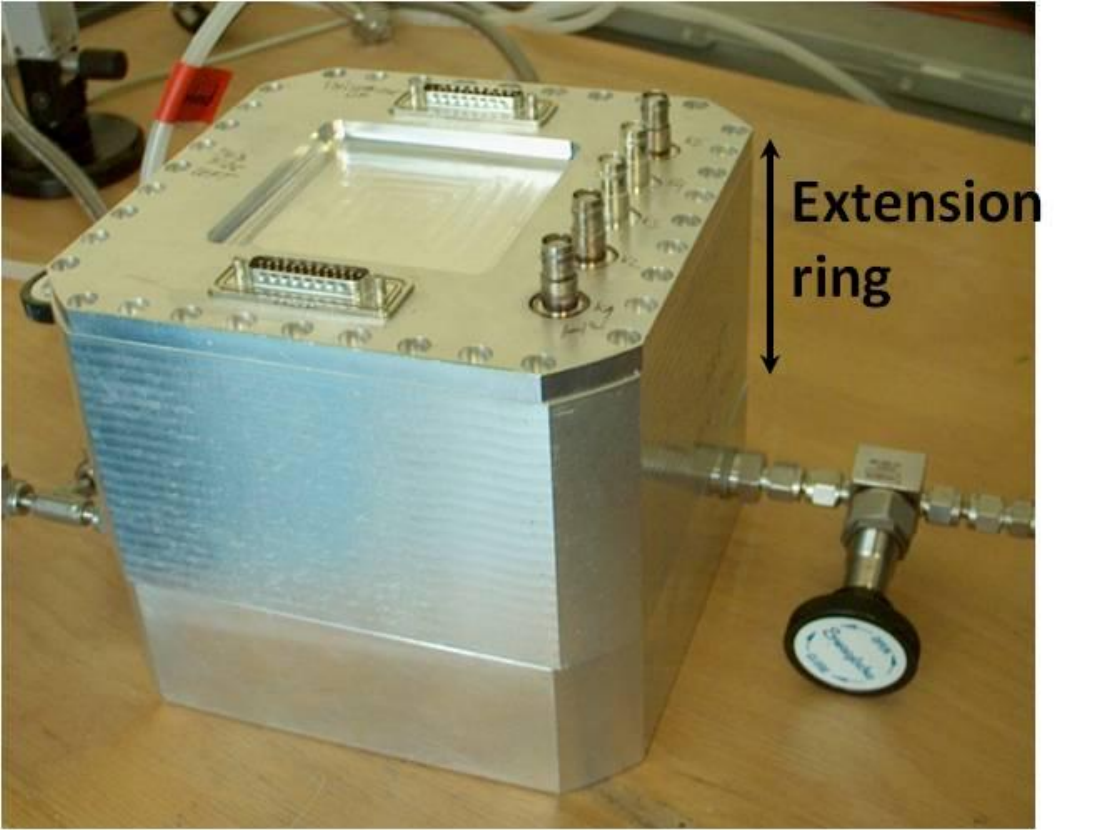}
\includegraphics[scale=0.6]{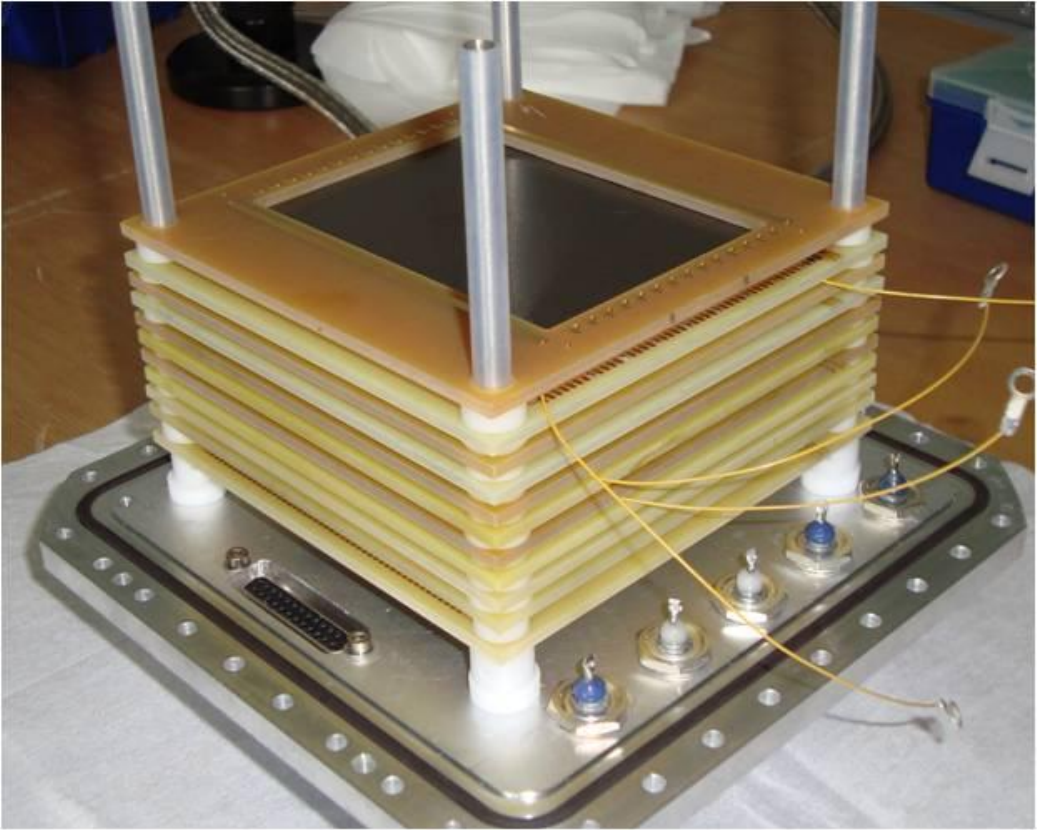}
\caption{Left: original test detector plus the extension ring used to measure the efficiency of the stack of five MPWCs. Right: stack incorporating five MWPCs with coated flat cathodes.}  
\label{v2}
\end{figure*}

As the dimensions of the original detector allowed to accommodate only up to four grooved plates and two wire planes integrated between them,  an extension ring  was fabricated and added to it in order to increase its height. This made efficiency measurements for stacks of up to five MPWCs possible, see Fig. \ref{v2}.  Each MWPC contains two Boron layers deposited on the surfaces facing the wire plane. Thus, the measurement with the stack of five counters was used to extract information concerning the efficiency of a number of 10 Boron layers arranged in an alternating  forward-backward irradiation geometry. The signal from each individual counter was read out separately by a MESYTEC preamplifier and fed to the multichannel analyzer. 

\begin{figure*}[ht]
\centering
\includegraphics[scale=0.35]{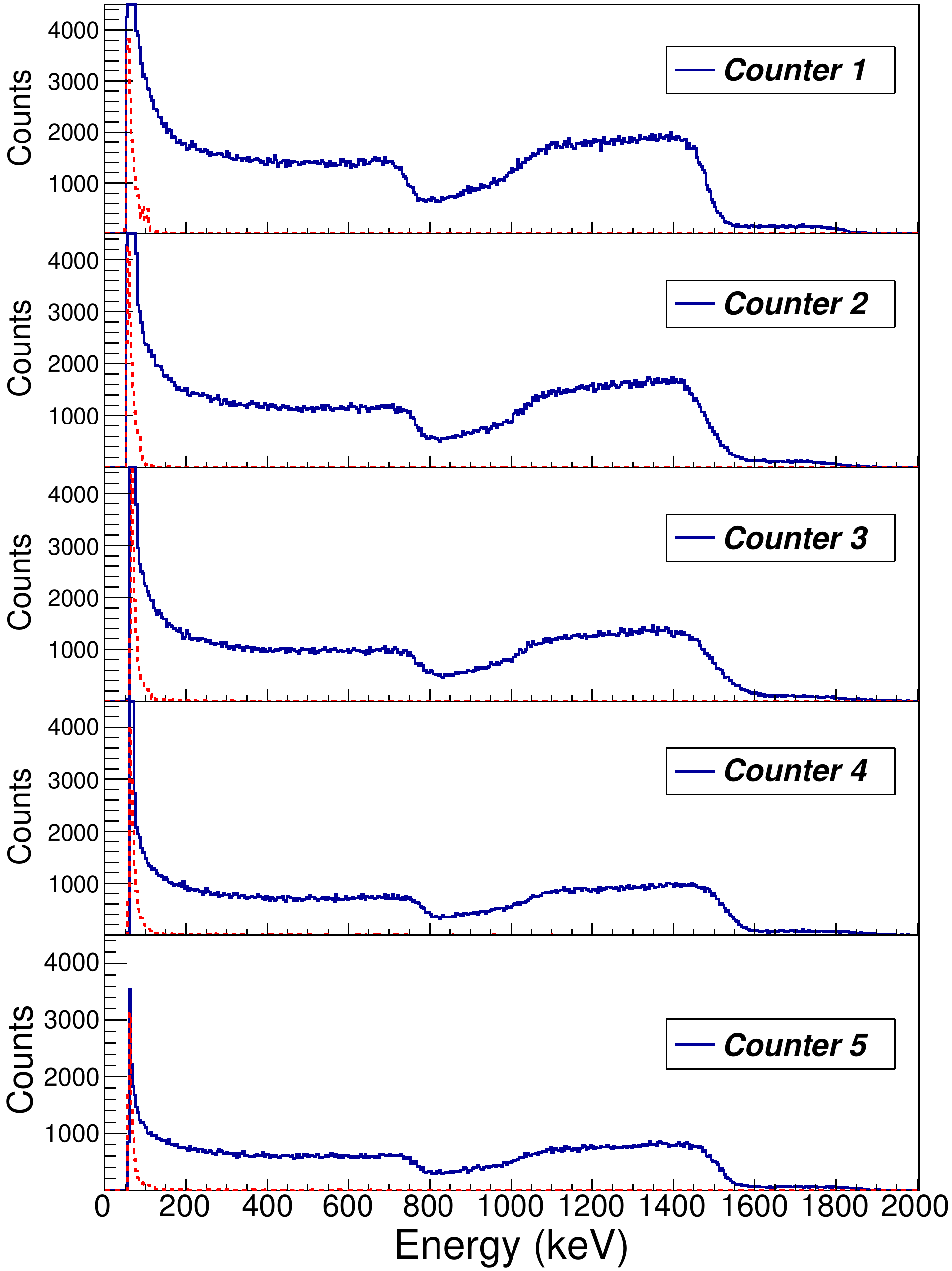}
\includegraphics[scale=0.35]{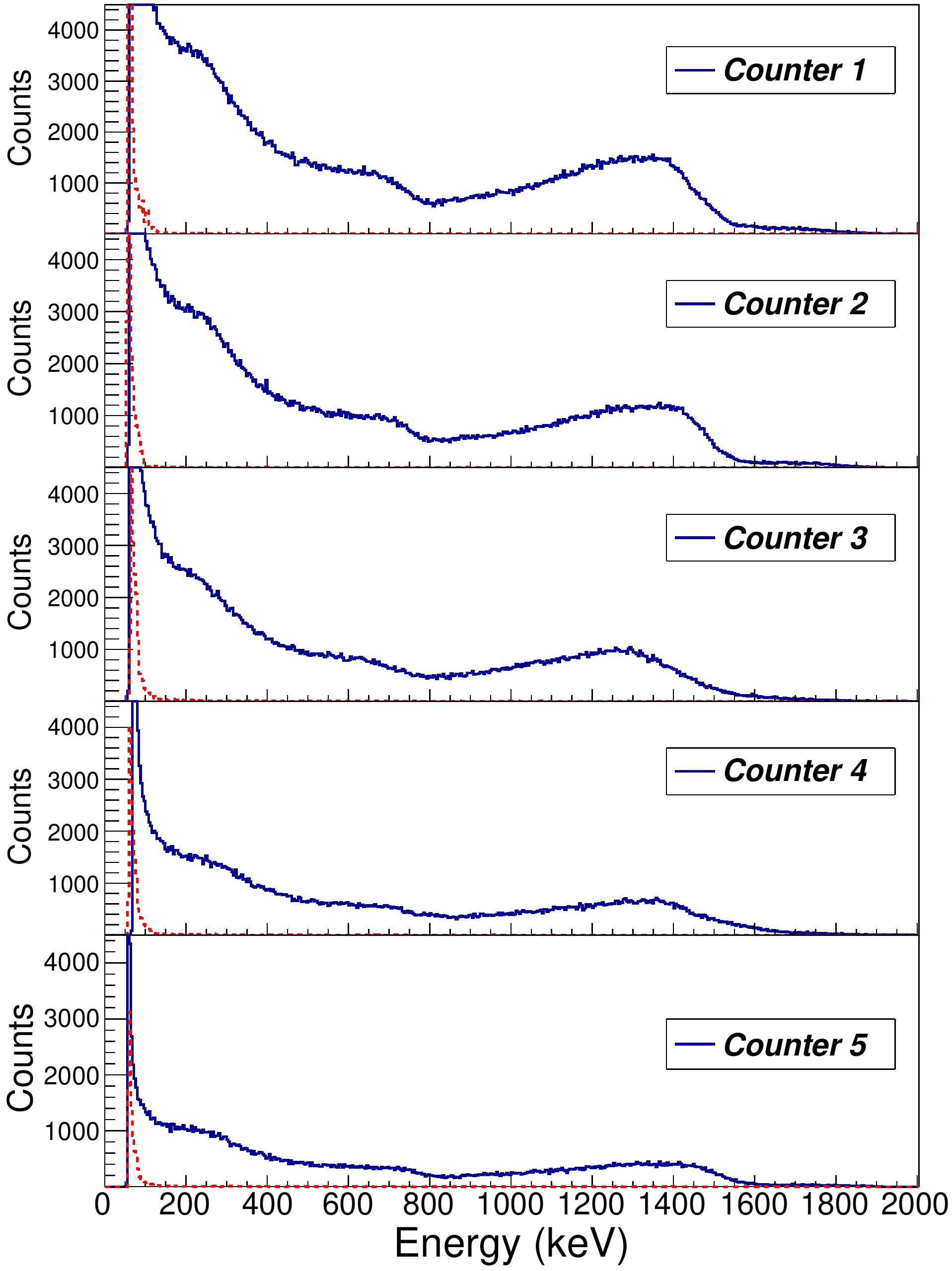}
\caption{Measured pulse-height spectra with MPWCs with coated parallel plates with flat (left panel) and macrostructured surface (right panel).  The thickness of the Boron layers was 0.95 $\mu$m on the flat plates and 1.1  $\mu$m on the grooved ones. All spectra were represented on the same ``Y" scale in order to facilitate the comparison. The spectra represented in red were taken with the neutron beam switched off and contain the signal generated by the environmental neutron and $\gamma$-rays as well as the electronics noise.  }  
\label{mult_sp}
\end{figure*}

The pulse-height spectra recorded by each individual counter of the stack is illustrated in Fig. \ref{mult_sp}. The left panel of the figure shows the spectra collected with the counters incorporating 0.95 $\mu$m thin Boron layers coated on 0.5 mm flat Aluminum-substrates. The top spectrum corresponds to the counter which is first hit by the incoming neutron beam and therefore, has the highest statistics. As the neutron beam passes through the detector and  interacts with the Boron layers, the flux is attenuated exponentially, leading to a decrease in the observed countrate. Apart from the decrease in the statistics with increasing the counter number, all pulse-height spectra look very similar, featuring the broad  distributions corresponding to the 1.47-MeV $\alpha$-particle and 0.84-MeV $^7$Li ion from the most intense $^{10}$B(n,$\alpha$)$^7$Li reaction channel \cite{Ste01}.  At energies below 100 keV, an exponential increase in statistics is observed. This signal arises from the reaction products losing most of their kinetic energy in the converter and  $\gamma$-background.  

The spectra displayed in the right panel of Fig. \ref{mult_sp} were collected with stacks of counters incorporating the macrostructured cathodes coated with 1.1 $\mu$m of $^{10}$B$_4$C. An additional feature around 300 keV is present in all spectra. As already discussed in Ref. \cite{Ste01}, the events in this low-energy bump are generated in the reaction of the incident neutron with the coating deposited on the sidewalls of the groove, which have only a limited flight-path through the counting gas and therefore, give rise to a low-energy signal.     

\begin{figure*}[ht]
\centering
\includegraphics[scale=0.6]{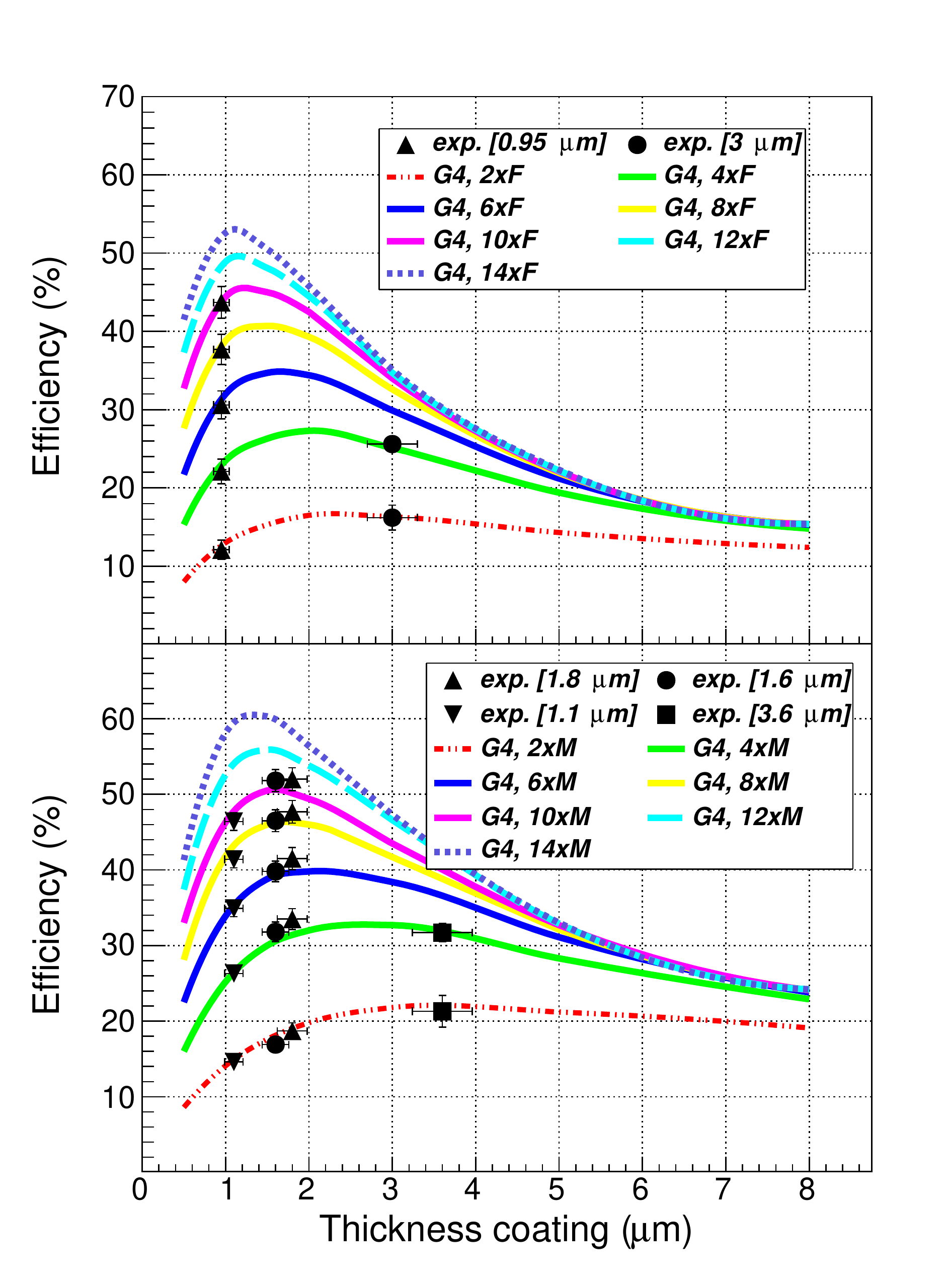}
\caption{Calculated and measured efficiencies for stacks of up to 10 Boron layers with various thicknesses coated on Aluminum-substrates with a flat surface (top) and double-sided macrostructured surface (bottom). The efficiencies were extracted from the calculated and measured pulse-height spectra, respectively, with a threshold set at 120 keV. The wavelength of the incident neutron beam was 4.7  \r{A}.}  
\label{eff_m}
\end{figure*}

Figure  \ref{eff_m} displays the measured and calculated efficiencies for stacks of up to five MWPCs with coated cathodes. The experimental efficiencies were extracted from the pulse-height spectra shown in Fig. \ref{mult_sp} with a threshold for $\gamma$-rejection set at an energy of 120 keV. The efficiency values for stacks of two or more counters were determined from the summed signal detected in the respective counters.  

As it can be seen in both panels of Fig. \ref{eff_m}, regardless of the surface topology, the distance between the efficiency curves calculated for the stacks with different numbers of layers becomes smaller with increasing the number of layers in the stack, indicating a non-linear accumulation of the total efficiency due to the attenuation of the neutron beam in the previous layers. Reaching detection efficiencies higher than 50$\%$ for 4.7 \r{A} neutrons requires 14 flat Boron layers coated on both sides with $\sim$1 $\mu$m and arranged in a 7-counter geometry. Obviously, this number becomes greater with decreasing the wavelength of the scattered neutrons to be detected. The calculations also indicate that for both types of surfaces,  the optimal thickness of the coating becomes smaller with increasing the number of layers in the stack, as expected \cite{Kle00,Pis13}.     

The increase in efficiency of a single Boron layer achieved by macrostructuring with grooves the surface of the substrate material upon which the neutron converter is deposited leads to an increase of the total efficiency of the detector incorporating stacks of macrostructured layers, when compared to a stack comprising the same number of layers, but with a flat surface. According to the experimental results shown in the bottom panel of Fig. \ref{eff_m}, only 10 coated  macrostructured layers mounted in 5 counters are needed to reach $\sim$50$\%$ detection efficiency for neutrons of 4.7 \r{A}. As mentioned above, the detector using the traditional flat surfaces for the cathodes requires a 7-counter setup in order to reach the same efficiency. This is an increase by $\sim$40$\%$ of the number of counters when compared to the stack of grooved layers. As any layer in the detector stack requires  detection of the ionization resulting from the reaction products that escape the Boron coating, the total number of the Boron layers employed to achieve the desired detection efficiency  is proportional to the number of readout structures (e.g., anode wire planes, pad-based modules, etc.). This leads to a large number of electronics channels, which greatly affects the complexity and price of the detector, especially for applications that require the coverage of square meters of active area.  The macrostructuring of the surfaces upon which the Boron layer is being deposited in order to increase the detection efficiency per counter provides a way to reduce the production and operation costs of the detector, without altering  the detection efficiency. 

\section{Conclusions}
\label{conc}

In the present work we report on the results of the evaluation of the performance of a detector based on the MWPC technology incorporating coated grooved-cathodes, a novel design that we have introduced in a previous paper \cite{Ste01}. The cathode was made by extrusion and consists of V-shaped grooves with an opening angle of 45$^\circ$ and a depth of $\sim$2 mm. A thin layer of $^{10}$B was deposited onto the surface of the grooves by employing DC magnetron sputtering. The GEANT4 model predicts that the efficiency of such cathode is $\sim$40$\%$ higher than the efficiency of a flat cathode of the same active area, owing to the larger capture probability for the incoming neutron by the converter deposited onto the sidewalls of the grooves. 

We tested experimentally the predictions of the GEANT4 model for the efficiency of a single Boron layer and found that the experimental and calculated data agree well with each other. Furthermore, we compared the measured and calculated efficiencies of stacks of up to 10 layers coated on grooved and flat surfaces, each group being arranged into 5 independent MWPCs enclosed in a common housing. The comparison shows that the efficiency of the stack of grooved cathodes is superior to that of a stack incorporating the same number of flat cathodes.  Our results indicate that the stack of layers with flat surfaces must incorporate $\sim$40$\%$ more layers in order to reach the efficiency of the detector comprising Boron layers deposited on grooved surfaces.  Thus, the use of coated grooved-cathodes in a neutron detector based on the MWPC technology allows one to obtain the desired detection efficiency with a reduced number of Boron layers. This translates into a reduction of the number of stacked counters  and consequently, number of electronics channels for the detector, while preserving the detection efficiency. Minimizing the production cost is a key factor for any alternative technology developed to replace the $^3$He-tubes in instruments which require large-area position-sensitive neutron detectors.  

The present R$\&$D work is of highest relevance for the design and construction of a $^3$He-free neutron detector with a performance to fully exploit the potential of the future European Spallation Source \cite{ess,ess-cdr}.  The conclusions of the work reported here are also relevant to other applications beyond those of thermal neutron scattering. Note also that whilst the MWPC was used here as the detecting element, the effect of the macrostructured converter substrates should be generally applicable; and that the MWPC could thus be replaced by an alternative suitable detecting element. 

\acknowledgments

This research was carried out in the framework of the German In-Kind Contribution to the ESS Design Update, Work Package DE-K2, and supported by the German Federal Ministry of Education and Research (BMBF) under Contract no. 05E10WO1.

\end{document}